\documentstyle[amssymb,prb,aps,epsf]{revtex}
\begin{document}
\draft

\advance\textheight by 0.5in \advance\topmargin by -0.in

\twocolumn[\hsize\textwidth\columnwidth\hsize\csname@twocolumnfalse\endcsname{ }

\title{ Schottky barriers in carbon nanotube heterojunctions}
\author{Arkadi A. Odintsov$^{1,2}$}
\address{
$^{1}$Department of Applied Physics and DIMES, Delft University of Technology, 
2628 CJ Delft, The Netherlands 
\\
$^{2}$Nuclear Physics Institute, Moscow State University, 
Moscow 119899 GSP, Russia }
\date{\today}
\maketitle

\begin{abstract}
Electronic properties of heterojunctions 
between metallic and semiconducting single-wall carbon nanotubes
are investigated.
Ineffective screening of the long range Coulomb interaction 
in one-dimensional nanotube systems
drastically modifies the charge transfer phenomena 
compared to conventional semiconductor heterostructures. 
The length of depletion region varies over a wide range
(from the nanotube radius to the nanotube length) 
sensitively depending on the doping strength.
The Schottky barrier gives rise to an asymmetry of the I-V characteristics 
of heterojunctions, in agreement with recent
experimental results by  Yao {\it et al.} and Fuhrer {\it et al.}
Dynamic charge build-up near the junction 
results in a step-like growth of the current at reverse bias.
\end{abstract}

\vskip -0.5 truein
]

Single-wall carbon nanotubes (SWNTs) are giant linear fullerene molecules
which can be studied individually by methods of nanophysics \cite{Dekker}.
Depending on the wrapping of a graphene sheet, SWNTs can either be
one-dimensional (1D) metals or semiconductors with the energy gap in
sub-electronvolt range \cite{Wildoer,Odom}. While metallic nanotubes can
play a role of interconnects in future electronic circuits, their
semiconducting counterparts can be used as basic elements of switching
devices. An example is the field effect transistor on semiconducting SWNT
operating at room temperature \cite{Tanstubefet}.

Of particular interest are all-nanotube devices \cite{Dresselhaus}. The
simplest can be fabricated by contacting two SWNTs with different electronic
properties. The SWNTs can be seamlessly joined together by introducing
topological defects (pentagon-heptagon pairs) into the hexagonal graphene
network \cite{Dunlap}. The resulting on-tube junction generically has the
shape of a kink. Electronic properties of such junctions have been
investigated theoretically (see e.g. Refs. \cite{Lambin,Chico}) within the
model of non-interacting electrons.

Electron transport in nanotube heterojunctions has been studied in two
recent experiments. Yao {\it et al.} treated junctions in SWNTs with kinks 
\cite{Yao} whereas Fuhrer {\it et al.} explored contacts of crossed
nanotubes \cite{McEuen}. Both groups observed non-linear and asymmetric $I-V$
characteristics resembling that of rectifying diodes. On one hand, the
rectifying behavior can be naturally interpreted in terms of Schottky
barriers (SBs). On the other hand, formation of a SB might be surprising
since one expects no charge transfer in junctions between two SWNTs made of
the same material.

A possible reason for the charge transfer might be the doping of the
nanotubes forming the heterojunction \cite{Farajian}. The doping can be
caused by introduction of dopant atoms into the nanotubes or by charge
transfer from metallic electrodes. In the latter case the doping strength
can also be controlled by the gate voltage. It is important to mention that
screening of the Coulomb interaction is ineffective in one-dimensional
nanotubes. For this reason the effect of the doping is long-ranged:
the density of the transferred charge decays slowly with the distance from
the electrodes and might be appreciable at the heterojunction \cite
{OdintsovTokura}.

The long-range Coulomb interaction should be properly taken into account
when treating the charge transfer in the heterojunction itself.
Unfortunately, this was not accomplished in Ref. \cite{Farajian}, where the
electric field was assumed to be fully screened in the region of a few
atomic layers near the junction. In this Letter we study charge transfer
phenomena in nanotube heterojunctions with true long-range Coulomb
interaction. We concentrate on the metal-semiconductor SWNT junction and
analyze its equilibrium and non-equilibrium properties (SB parameters, I-V
characteristics) by solving the Poisson equation self-consistently.

As a model system we consider ''straight'' junction \cite{Pi} between
metallic ($x<0$) and semiconducting ($x>0$) SWNTs (Fig. 1). We assume that
the conducting $p_{z}$ electrons in SWNTs are confined to the surface of a
cylinder of radius $R$. The nanotubes are surrounded by a coaxial
cylindrical gate electrode of radius $R_{s}\gg R$. The Fourier components of
the 1D Coulomb interaction are given by 
\begin{equation}
U(q)=\frac{2e^{2}}{\kappa }\left\{ I_{0}(qR)K_{0}(qR)-\frac{%
I_{0}^{2}(qR)K_{0}(qR_{s})}{I_{0}(qR_{s})}\right\} ,  \label{U}
\end{equation}
with the dielectric constant of the medium $\kappa $ and the modified Bessel
functions $I_{0}$, $K_{0}$. Equation (\ref{U}) describes the long-range
Coulomb interaction, $U(x)=1/\kappa x$, for $R\ll x\ll R_{s}$. The
interaction is screened at large distances $x\gg R_{s}$, so that $%
U(0)=e^{2}/C=(2e^{2}/\kappa )\ln (R_{s}/R)$, $C$ being the capacitance of
SWNT per unit length.\ The kernel (\ref{U}) relates the electrostatic
potential $\varphi $ at the surface of SWNTs to 1D charge density $e\rho $ ($%
e>0$), 
\begin{equation}
e\varphi _{q}=U(q)\rho _{q}.  \label{phi(q)}
\end{equation}
\begin{figure}[tbp]
\hspace{0.1\columnwidth}\epsfxsize=0.8\columnwidth\epsfbox{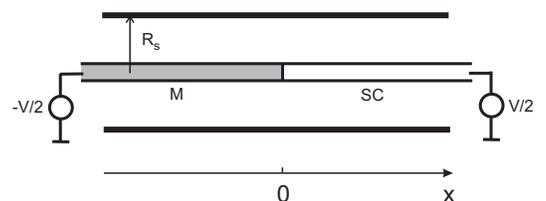}
\caption{Heterojunction between metallic (M) and semiconducting (SC)
nanotubes. The potential $V_{g}$ is applied to a cylindrical gate electrode
of radius $R_{s}$}
\label{fig1}
\end{figure}

Since experimental values \cite{Yao,McEuen} of the conductance of
heterojunctions are small, $G/(e^{2}/h)\lesssim 10^{-2}$, we will assume low
transparency $T\ll 1$ of the barrier between two SWNTs. In this case the
electrons in the nanotubes are described by the equilibrium Fermi
distribution $f(E)$, also when the voltage $V$\ is applied to the system.

In equilibrium, the charge density is related to the energy $\tilde{E}%
_{0}(x)=E_{0}(x)-E_{F}(x)$ of the gapless point (charge neutrality level) of
graphite $E_{0}$ counted from the Fermi level $E_{F}$,

\begin{equation}
\rho (x)=\int dEsign(E)\nu (E)f[(E-\tilde{E}_{0}(x))sign(E)],  \label{rho(x)}
\end{equation}
with the density of electronic states $\nu $ \cite{Slowvar}. Equation (\ref
{rho(x)}) is valid provided that $\tilde{E}_{0}(x)$ varies slowly on the
scale of the Fermi wavelength.

We restrict our consideration to low energies $|\tilde{E}_{0}|<\Delta ^{(1)}$%
, $k_{B}T\ll \Delta ^{(1)}$ and neglect the effect of higher 1D subbands ($%
\Delta ^{(1)}/(\hbar v_{F}/R)=1,2/3$ for metallic/semiconducting SWNT). The
densities of states in metallic and semiconducting SWNTs are given by 
\begin{equation}
\nu _{M}=\frac{4}{\pi \hbar v_{F}},\ \nu _{S}=\frac{4}{\pi \hbar v_{F}}\frac{%
|E|\Theta (|E|-\Delta )}{\sqrt{E^{2}-\Delta ^{2}}},  \label{dos}
\end{equation}
with the Fermi velocity $v_{F}\simeq 8.1\times 10^{5}$ m/s and the energy
gap $2\Delta =2\hbar v_{F}/3R$ in semiconducting SWNT ($\Delta \simeq 0.3$
eV for generic SWNTs \cite{Wildoer} with $R=0.5-0.7$ nm).

In the limit of zero temperature Eq. (\ref{rho(x)}) may be inverted as, 
\begin{equation}
\tilde{E}_{0}(\rho )=\left\{ 
\begin{array}{c}
\rho /\nu _{M},\;x<0, \\ 
\sqrt{\Delta ^{2}+\left( \rho /\nu _{M}\right) ^{2}},\;x>0.
\end{array}
\right.  \label{Krho}
\end{equation}
The charge neutrality level $\tilde{E}_{0}(x)$ is related to the
electrostatic potential (\ref{phi(q)}), 
\begin{equation}
\tilde{E}_{0}(x)+e\varphi (x)=\mu +eVsign(x)/2,  \label{Eqmu}
\end{equation}
$\mu \mp eV/2$ being the electro-chemical potentials for holes in metallic
and semiconducting SWNTs. The potential $\mu =\alpha (\Delta W-eV_{g})$ can
be controlled by the gate voltage $V_{g}$ (Fig. 1). It also incorporates the
difference $\Delta W=W_{M}-W_{NT}$ of the work functions of the gate
electrode and SWNT \cite{Landau} (the coefficient $\alpha $ characterizes
mutual capacitance of the nanotubes to the gate and is equal to unity in our
case).

We solve Eqs. (\ref{phi(q)}), (\ref{rho(x)}), (\ref{Eqmu})
self-consistently\ by numerical minimization of the corresponding energy
functional. The Coulomb energy is computed in the Fourier space. Figures 2,
3 display the results for the following parameters: $R_{s}/R=75$ ($%
R_{s}\simeq 50$ nm for (10,10) SWNTs) and $\nu _{M}U(0)/\ln (R_{s}/R)=5$.
The latter value corresponds to the dielectric constant $\kappa \simeq 1.4$\
which can be inferred from the experimental data (see Fig. 4 of Ref. \cite
{Dekker}).

The band bending diagrams (Fig. 2) display the charge neutrality level $\bar{%
E}_{0}(x)=\tilde{E}_{0}(x)-eVsign(x)/2$ counted 
from the ''average'' Fermi level of metallic and semiconducting
SWNTs, as well as the energies $E_{c,v}=\bar{E}_{0}\pm \Delta $ 
\begin{figure}[tbp]
\epsfxsize=\columnwidth\epsfbox{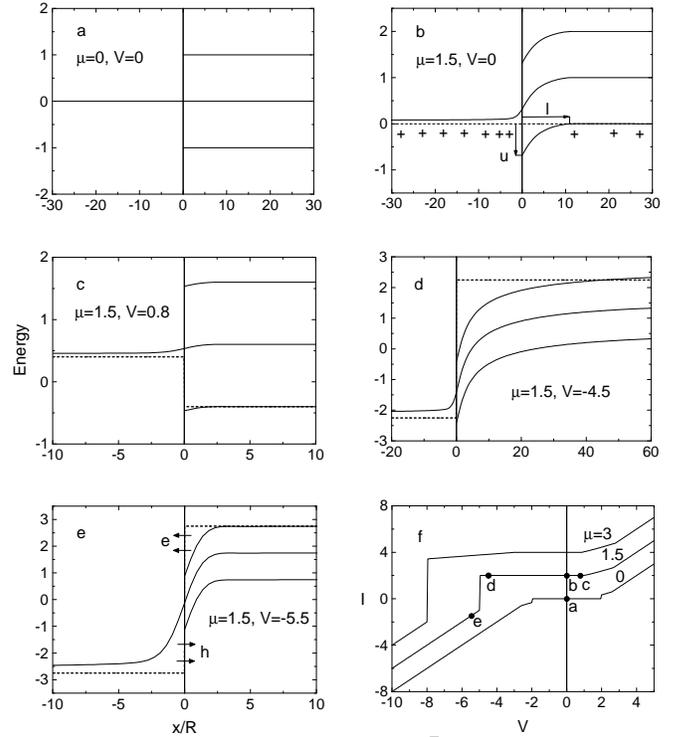}
\caption{The charge neutrality level $\bar{E}_{0}$ and the energies of the
conduction $E_{c}$ and valence $E_{v}$ bands as functions of the distance
from the junction (a-e). The Fermi levels are shown by dashed lines. The $%
I-V $ characteristics of the heterojunction at zero temperature (f). The
energies $\protect\mu $, $eV$ are in units of $\Delta $; the current is in
units of $2e\Delta T_{i}/(\protect\pi \hbar )$. The $I-V$ curves for $%
\protect\mu =1.5,3$ are offset for clarity. }
\label{fig2}
\end{figure}
\noindent of the
conduction and valence bands in semiconducting SWNT. Let us start from the
case of zero bias, $V=0$ (Figs. 2(a), 2(b)). At zero electro-chemical
potential, $\mu =0$, the Fermi level of the nanotubes coincides with the
gapless point of graphite and the system is charge neutral (Fig. 2(a)). This
situation occurs for isolated nanotubes. The barriers for the electron and
hole transport are equal to $\Delta $ (Fig. 3(a)).

To make contact with the experiments \cite{Yao,McEuen} we will concentrate
on p-doped SWNTs ($\mu >0$). Due to larger number of electronic states $%
\int\nolimits_{0}^{\tilde{E}_{0}}dE\nu (E)$ the metallic SWNT acquires more
charge and has higher electrostatic potential $\varphi (-\infty )$ (lower
charge neutrality level $\tilde{E}_{0}(-\infty )$) compared to
semiconducting SWNT kept at the same electrochemical potential, see Eqs. (%
\ref{Krho}), (\ref{Eqmu}). The electric field induced by this charge bends
the bands in the semiconducting part downwards so that a SB is formed near
the interface (Fig. 2(b)).

For $|\mu |<\Delta $, there are no free charges in the semiconducting SWNT.
Our numerical results indicate that the electrostatic potential $\varphi (x)$
decays logarithmically at $R\ll x\ll R_{s}$ so that the bend bending
extends over long distances $x\sim R_{s}$ (the analytical estimate, $%
\varphi (x)\simeq e\nu _{M}\mu \ln (R_{s}/x)/\kappa $, is available in the
limit of weak interaction, $\nu _{M}U(0)\ll 1$). 
At $\mu =\Delta $ holes enter the
semiconducting SWNT. With increasing the electro-chemical potential the
holes come closer to the junction reducing the length $l$ and the height $u$
of a SB (Fig. 2(b)). In the case of weakly doped semiconducting SWNT, $\mu
=\Delta (1+\delta )$, $\delta \ll 1$, a rough estimate of the depletion
length $l$ can be made, $\ln (l/R_{s})\sim \delta \ln (R/R_{s})$, for $R\ll
l\ll R_{s}$. Therefore, the depletion length changes rapidly from $l\sim
R_{s}$ to $l\sim R$ with increasing doping in this regime. The height of a
SB can be estimated from the difference of the charge neutrality levels in
semiconducting and metallic SWNTs, $u\lesssim \tilde{E}_{0}(\infty )-\tilde{E%
}_{0}(-\infty )$. The latter evaluate at $\tilde{E}_{0}(-\infty )=\mu
/(1+\nu _{M}U(0))$ and $\tilde{E}_{0}(\infty )=\Delta $ for $\delta \ll \nu
_{M}U(0)$, see Eqs. (\ref{phi(q)}), (\ref{Krho}), (\ref{Eqmu}). Since the
band bending occurs predominantly in the semiconducrting part (Figs. 2(a),
2(b)) and $\nu _{M}U(0)\gg 1$,\ one expects that $u\simeq $ $\Delta $ for $%
\delta \ll \nu _{M}U(0)$. Note that SB persists up to remarkably large
values of the electro-chemical potential, $\mu \approx 14\Delta $ (Fig.
3(a)) though it becomes rather short ($l<R$) for $\mu \gtrsim 8\Delta $.

Figure 3(a) shows the result for the SB height defined as the minimum energy
of electron or hole excitation required to transfer the elementary charge
across the junction in the absence of tunneling through the SB. The SB
height shows pronounced asymmetry as a function of the bias voltage. Under a
forward/backward bias the charge density in metallic SWNT
decreases/increases. This reduces/enhances the band bending (Figs. 2(c),
2(d)) in the semiconducting part (the charge density and the band bending
change sign for $V>2\mu $, cf. Fig. 3(a)). As a result, the SB height
decreases faster under a forward bias. This gives rise to the asymmetry $%
V_{+}<|V_{-}|$ of positive $V_{+}$ and negative $V_{-}$ threshold voltages
at which SB vanishes ($u\rightarrow 0$) and the onset of the conductance
occurs. The positive threshold voltage is relatively insensitive to the
doping strength, Fig. 3(a). This can be used for a rough estimate of the gap 
$\Delta $ from experimental data, $\Delta \simeq eV_{+}$, for $\mu \sim 1$.
Note that we assume weak interband tunneling in semiconducting SWNT so that
the electronic states in the conductance band are empty in Fig. 2(d).

We will proceed with the analysis of non-equilibrium electron transport. The
current through the heterojunction is given by the Landauer formula, 
\begin{equation}
I=\frac{2e}{\pi \hbar }\int dET(E)\left\{ f(E-eV/2)-f(E+eV/2)\right\} ,
\label{Current}
\end{equation}
with the energy-dependent transmission coefficient $T(E)$ of the junction.
It is natural to separate the contribution $T_{i}(E)$ of a barrier at the
interface between SWNTs \cite{Chico} and the contribution $T_{S}(E)$ of a
SB to the total transmission. As a minimal model, we assume that the
transparency $T_{i}$ is energy independent whereas the transparency $T_{S}(E)
$ increases from zero to unity when the energy $E$ crosses the edge of a SB.
In this case the total transmission reads $T(E)=0,(T_{i})$, for the energies
in (out of) the SB range $[E_{\min },E_{\max }]$. In the case of downward
bending (Fig. 2) the SB range is given by $[E_{v}(0),E_{c}(\infty )]$,
in the absence of charge carriers in the conduction band, $\mu +eV/2>-\Delta 
$ (Figs. 2(a-d)), and by $[E_{v}(0),E_{c}(0)]$, in their presence, $\mu
+eV/2<-\Delta $ (Fig. 2(e)). The results for the $I-V$ characteristics at
zero temperature are presented in Figs. 2(f), 3(b).

With increasing forward bias the SB (Fig. 2(c)) disappears and the hole
transport channel opens at $V>V_{+}$. 
\begin{figure}[tbp]
\hspace{0.1\columnwidth}\epsfxsize=0.75\columnwidth\epsfbox{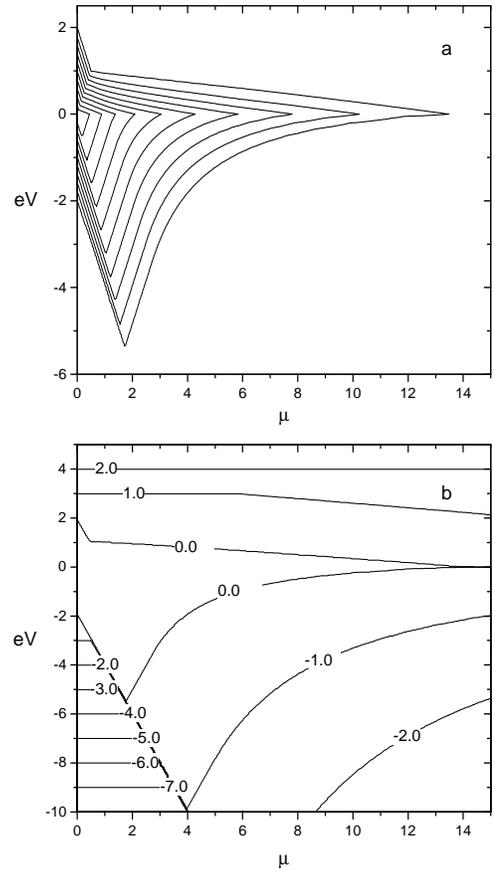}
\caption{The height $u$ of the Schottky barrier (a) and the current through
the heterojunction (b) at zero temperature. The contour lines (a) correspond
to $u/\Delta=0, 0.1,..., 0.9$ from periphery to origin. The energies $%
\protect\mu$, $eV$ are in units of $\Delta$. The current is in units of $%
2e\Delta T_{i}/(\protect\pi\hbar)$. }
\label{fig3}
\end{figure}
\noindent The cusp at the $I-V$ characteristics (Fig. 2(f)) at somewhat
higher voltages corresponds to the onset of the electron channel. Note that
at high (forward or reverse) bias both the electron and hole channels are
open and the current is given by $I=(2eT_{i}/\pi \hbar )[eV-2\Delta sign(V)]$%
.

The onset of the conductance under reverse bias depends critically on the
electro-chemical potential $\mu $ (Figs. 2(f), 3(b)). At low doping, $\mu
\lesssim 1.8\Delta $, the current increases abruptly at $V_{-}=V_{c}$, $%
eV_{c}=-2(\Delta +\mu )$. The voltage $V_{c}$ corresponds to the alignment
of the Fermi level with the conduction band of semiconducting SWNT.
Electrons entering the conduction band cause the charge build-up near the
junction. The reconstruction of the band profile (cf. Figs. 2(d) and 2(e))
results in the onset of the electron and hole channels of transport giving
rise to a step-like growth of the current. At higher doping, $\mu \gtrsim
1.8\Delta $, the threshold voltage $V_{-}>V_{c}$ corresponds to the opening
of the hole channel. The current gradually grows under reverse bias $%
V_{c}<V<V_{-}$ until the reconstruction of the band profile occurs at $%
V=V_{c}$ (see the curve for $\mu =3$ in Fig. 2(f)).

We now consider quantum tunneling through the SB. The transparency $T_{S}(E)$
of the SB can be evaluated using WKB method and the effective mass
approximation. For a triangular barrier of the length $l$ and the height $u$
we obtain, 
\begin{equation}
T_{S}\sim \exp \left( -\frac{4l}{9R}\sqrt{\frac{2u}{\Delta }}\right) .
\label{Ts}
\end{equation}
The transparency $T_{S}$ increases considerably near the boundaries of the
transport blockade region $[V_{-},V_{+}]$ (Fig. 3(a)) due to decreasing $u$
and $l$. For example, $T_{S}\sim 2.5\times 10^{-3}$ for the SB in Fig. 2(b),
whereas $T_{S}\sim 1$ for the SBs in Figs. 2(c), 2(d). This gives rise to a
substantial leakage current in the blockade region.

The asymmetry of the $I-V$ characteristics and threshold voltages has been
discovered in recent experiments \cite{Yao,McEuen}. According to the data of
Ref. \cite{Yao}, both the thresholds $V_{+}$, $V_{-}$ shift upwards with the
gate voltage. Moreover, the positive threshold shifts less than the negative
one. Such behavior is consistent with our model in the regime of moderate
doping, $0.5<\mu /\Delta \lesssim 1.8$ (Fig. 3). However, the blockade
region of $3-4$ V detected in the experiment is somewhat wider than the
theoretical estimate, $V_{+}-V_{-}\lesssim 6.5\Delta \simeq 2$ eV. The extra
voltage drop could be due to potential disorder in semiconducting SWNT \cite
{McEuenDisorder} and/or an additional SB at the interface between
semiconducting SWNT and metallic electrode.

We now check the model against the experimental data of Ref. \cite{McEuen}.
The measured width of the blockade region, $0.5-0.7$ V, agrees with the
theoretical estimate. The gap in semiconducting SWNT, $\Delta \simeq eV_{+}$%
, evaluates at $\Delta =0.19$, $0.29$ eV for the two devices studied \cite
{McEuen}. These values are in the expected range $\Delta \sim 0.25-0.35$ eV 
\cite{Wildoer,Odom}. A smooth onset of the current over the range $\sim
0.1-0.3$ eV around threshold voltages is naturally associated with quantum
tunneling through a ''leaky'' SB (thermal energies are much smaller, $%
k_{B}T\simeq 5$ meV). Finally, the step-like feature of the current under
reverse bias almost certainly corresponds to the reconstruction of the band
profile due to the Fermi level entering the conduction band of
semiconducting SWNT. Gradual onset of the differential conductance following
the reconstruction might be associated with increasing conductance of
disordered semiconducting SWNT under the doping \cite{McEuenDisorder}.

To conclude, we have studied the electronic properties of carbon nanotube
heterojunctions and provided explanation for the main features of recent
experimental data \cite{Yao,McEuen}. Due to the long-range Coulomb
interaction, the charge transfer phenomena in one-dimensional nanotube
systems differ drastically from those in conventional semiconductor
heterostructures. This creates new challenges in the design of novel
electronic devices. In particular, the long-range electrostatic potential in
underdoped junctions might affect other components of a circuit, whereas
substantial leakage current in overdoped junctions spoils the rectification.
In view of these challenges a new concept of functional devices on molecular
level might be needed.

In the process of writing this paper I became aware of the preprint by
L\'{e}onard and Tersoff \cite{Leonard} who investigated equilibrium
properties of junctions between semiconducting SWNTs and found the
long-range charge-transfer phenomena in these systems (see also Ref. \cite
{OdintsovTokura}).

The author wishes to thank B.L. Altshuler, G.E.W. Bauer, Yu.V. Nazarov, S.
Tarucha, Y. Tokura, Z.Yao, and, especially, C. Dekker and P. McEuen for
stimulating discussions. F. L\'{e}onard and J. Tersoff are acknowledged for
sharing the results of Ref. \cite{Leonard} before publication. This work was
supported by the Royal Dutch Academy of Sciences (KNAW).


\bigskip

\begin{references}
\bibitem{Dekker}  For a recent review see C. Dekker, Physics Today {\bf 5},
22 (1999).

\bibitem{Wildoer}  J.W.G. Wild\"{o}er, L.C. Venema, A.G. Rinzler, R.E.
Smalley, and C. Dekker{\it , } Nature {\bf 391}, 59 (1998).

\bibitem{Odom}  T.W. Odom, J. Huang, P. Kim, and C.M. Lieber, Nature {\bf 391%
}, 62 (1998).

\bibitem{Tanstubefet}  S.J. Tans, A.R.M. Verschueren, and C. Dekker, Nature 
{\bf 393}, 49 (1998).

\bibitem{Dresselhaus}  M. Dresselhaus, Physics World {\bf 5}, 18 (1996).

\bibitem{Dunlap}  B.I. Dunlap, Phys. Rev. B {\bf 49}, 5643 (1994).

\bibitem{Lambin}  Ph. Lambin, A. Fonseca, J.P. Vigneron, J.B. Nagy, and A.A.
Lucas, Chem. Phys. Lett. {\bf 245}, 85 (1995).

\bibitem{Chico}  L. Chico, L.X. Benedict, S.G. Louie, and M.L. Cohen, Phys.
Rev. B 54, 2600 (1996).

\bibitem{Yao}  Z. Yao, H. Postma, L. Balents, and C. Dekker, to be published
in Nature.

\bibitem{McEuen}  M.S. Fuhrer, J. Nyg\aa rd, L. Shih, M. Bockrath, A. Zettl,
and P. McEuen{\it ,} submitted to Nature.

\bibitem{Farajian}  A.A. Farajian, K. Esfarjani, and Y. Kawazoe, Phys. Rev.
Lett. {\bf 82}, 5084 (1999).

\bibitem{OdintsovTokura}  A.A. Odintsov, Y. Tokura, to be published in {\em %
Proceedings of the LT-22, Helsinki, 1999}, preprint cond-mat/9906269.

\bibitem{Pi}  We expect the results to be qualitatively correct for
heterojunctions with angles $\chi \gtrsim \pi /2$.

\bibitem{Slowvar}  In Eq. (\ref{rho(x)}) $\rho $ is averaged over few atomic
distances. Our approach does not describe phenomena at atomic lengthscale,
for instance, the Friedel oscillations.

\bibitem{Landau}  We assume that the charges at the nanotube and the gate
electrode are not compensated by e.g. atmosphere ions, see L.D. Landau and
E.M. Lifshitz, Electrodynamics of Continuous Media, Pergamon Press 1960, Ch.
3.

\bibitem{McEuenDisorder}  P.L. McEuen, M. Bockrath, D.H. Cobden, Y. Yoon,
and S.G. Louie, preprint cond-mat/9906055.

\bibitem{Leonard}  F. L\'{e}onard and J. Tersoff, preprint.
\end{references}
\end{document}